# IoT Solutions with Multi-Sensor Fusion and Signal-Image Encoding for Secure Data Transfer and Decision Making*

Piyush K. Sharma[1], Mark Dennison[2], and Adrienne Raglin[2]

Army Research Laboratory[1,2]
Adelphi, MD 20783, USA

**Abstract.** Deployment of Internet of Things (IoT) devices and *Data Fusion* techniques have gained popularity in public and government domains. This usually requires capturing and consolidating data from multiple sources. As datasets do not necessarily originate from identical sensors, fused data typically results in a complex data problem. Because military is investigating how heterogeneous IoT devices can aid processes and tasks, we investigate a multi-sensor approach. Moreover, we propose a signal to image encoding approach to transform information (signal) to integrate (fuse) data from IoT wearable devices to an image which is invertible and easier to visualize supporting decision making. Further-more, we investigate the challenge of enabling an intelligent identification and detection operation and demonstrate the feasibility of the proposed *Deep Learning* and *Anomaly Detection* models that can support future application that utilizes *hand gesture data* from wearable devices.

## 1 Introduction

Advances in technology have contributed to the growing use of laptops, smart-phones, and tablets in government and industry globally. Nascent 5G network enables mobile computing more effectively [13]. Increasing amount of information ow are impacting daily life of individuals, business conducted across industry, and government operations [3]. Expansion of digital information introduces several challenges related to complexity of data management, storage, privacy, processing and transfer. These challenges are also seen in utilizing information for decision making. In order to leverage this abundant, raw, digital information, military must keep abreast with advanced data-driven technology to meet mission requirements.

---
* This work was supported by Army Research Laboratory (ARL) Grant# W911NF-18-2-0043.

Modern military's Multi Domain operations (MDO) can potentially utilize Internet of Things (IoT) technology connecting land, water, air, space and cyberspace in a cohesive network [10,2,1]. This increases situational awareness and risk assessment, reduces response time, and can help prepare military for the future battle elds. IoT devices capture and consolidate data from multiple sources allowing the full realization of C2 (Command and Control) system to provide situational awareness [36]. This combined information gathered by a range of heterogeneous IoT devices can give an edge for potential strategic advantages.

IoT allows collection and dissemination of digital information by deploying several devices which output massive amount of data [38], [39]. Data related technologies provide IoT solutions to help industry and government to make better decisions by exploring underlying data regularities and revealing patterns, and provide an early warning to act upon possible threats (malware, terrorism, fraud, etc.) [4]. This allows military to take agile, perceptive, resilient, and re- liable decisions in timely manner to meet mission requirements. We seek out to address questions: How to tailor information for transmission over a com- promised network and retain most of it in the process with the goal of enabling decisions? How to model complex systems, detect and understand issues, and improve transmission?

Moreover, growing number of cyber-attacks put sensitive information at risk [32]. Information is a strategic asset for government organizations, and they need to protect it for national security [35]. Recently *Mirai botnet* malware launched a distributed denial of service (DDoS) attack causing much of the internet inaccessible on the U.S. east coast [11]. With the Army's growing dependency on IoT devices and the possible failure of current technology to keep up with the cyber security, C2 operations are susceptible to cyber-attack from possible threats capable of compromising IoT devices to exterminate multi-domain operations by injecting false or compromised information [18]. Hackers are developing new variants of IoT-focused malware with alarming regularity. Due to potentially sensitive nature of IoT datasets, *Blockchain technology* is used to facilitate se- cure sharing of IoT datasets, which allows digital information to be distributed, but not copied [26]. However, blockchain has several limitations related to complexity, scalability, and excessive energy consumption [27]. In order to leverage communication in C2 systems and to address the raised questions, we propose a sensor signal to image encoding approach to modify information (signal) by transforming it to an image which is invertible and easier to visualize [37].

Future military combat suits, helmets, weapons, and other equipment will be embedded with sensing and computing devices. These devices not only provide real-time interaction between soldiers and commanding officers in combat zones, but also help understand the psychophysical condition of soldiers. Modern devices are capable of collecting various aspects of biometric forensics, such as, data with facial, fingerprint, heartbeat, body temperature, etc. attributes. Consolidating data from disparate sensory devices can provide more consistent, accurate and relevant information helpful in understanding the state of mind and cognitive load during tasks.

One of the main goals of this paper is to investigate multi-sensor fusion approach producing combined data for improved accuracy and reduced uncertainty in channeled information [22]. A controlled exploitation of this approach can reduce cognitive load of soldiers, improve decision making, and establish a better soldier-machine (IoT devices) interaction; essential and ideal conditions for mission success. The purpose of proposed multi-sensor fusion and signal-image encoding approaches is to transform senseless and unstructured data into structured data from which meaningful information can be extracted to be used as actionable insights that enable intelligent military decision-making [36].

In Section 2, we provide background and related work. In Section 3, we define CJSD kernels and related research literature. In Section 4, we describe experimental data. In Section 5, we explain our signal-image encoding approach. In Section 6, we describe why we choose performed experiments. In Section 7, we provide a comparative analysis of results obtained with our experiments. Finally, we conclude our contribution with a summary of results in Section 8.

## 2   Background and Motivation

Potential of massive amount of data generated from a multi-sensor data fusion approach has limited value if it cannot be correlated with some context [25]. This plethora of information can be used in a context-based paradigm to improve personnel authentication accuracy using aforementioned biometric forensics. There can be added challenges to confirm identities of individuals in the fast pace environment that future soldiers might face. Because timing is crucial for mission success, suspicious and potential targets can be identified and neutralized instantly with Edge Computing [12].

The goal of this study is to measure the in influence different aspects of human state have on how a person makes gestures. These data were collected with the sole purpose of training machine learning algorithms to automatically detect the onset and o set of separate gestures in a series.

Research literature presents examples where multi-sensor data fusion techniques were implemented in different application domains. One paper proposed a smart home system with wearable intelligent technology, artificial intelligence, and multisensor data fusion technology [14]. A 3D gesture recognition algorithm was developed to recognize hand gestures. Another work presented a human activity recognition example using wearable and environmental sensor data fusion approach [29]. Furthermore, examples of Human-Computer-Interaction (HCI) for sign language and gesture recognition have been proposed within a multi- sensor fusion framework [19], [20].

Our work extends beyond the multi-sensor data fusion by introducing signal-image encoding scheme which transforms each fused data point into an image as described in Section 5. Research literature presents many examples of signal to image transformation techniques [45,15]. Our contribution is an information encoding approach which has multiple advantages; transformed data preserves allinformation, generated image les are small in size thus can easily be transferred over internet, transformed data is invertible and easier to visualize. Therefore, we can retrieve original signal from encoded image without requiring a key.

For image data obtained from signal transformation, we explore Transfer Learning approach which uses the pre-trained weights of a large image dataset [30]. This emerging approach has gained popularity in recent years and it intends to improve the performance of a neural network trained on a small dataset. For fusion sensor (signal) data, we explore recently introduced information theoretic kernels, Chisini Jensen Shannon Divergence (CJSD) and its metric version, Metric-Chisini Jensen Shannon Divergence (M-CJSD), known for their utility to tease apart data classes with discernible differences in Support Vector Machine (SVM) classification [43,41,40,42,38,39]. For benchmarking, we implement a multi-layered Deep Neural Network (DNN) model and tune it on a large hyperparameter grid. Furthermore, we explore a number of novelty detection techniques for their ability to identify anomalous observations which deviate significantly from the majority of data in contrast to classification models which assign predicted classes into groups. We perform a comparative analysis of employed models and measure their performances in terms of classification accuracy, sensitivity and specificity. Moreover, we compute confusion matrices for comparative analysis of our models. We evaluate all implemented methods on hand gesture datasets described in Section 4.

## 3   Family of CJSD and M-CJSD Kernels

Let, $P = \{p_i\}_{i=1}^N$, and $Q = \{q_i\}_{i=1}^N$ be two probability distributions, where $p_i$ and $q_i$ are the respective probabilities associated to the $i^{th}$ state (possible values). The family of CJSDs is defined in [43] with the following expression:

$$CJSD(P||Q) = \frac{1}{2}\left[\sum_{i=1}^N p_i \log\frac{p_i}{M_i} + \sum_{i=1}^N q_i \log\frac{q_i}{M_i}\right] \tag{1}$$

where, $M$ is the mid-point of distributions $p_i$ and $q_i$. If $M$ is the arithmetic mean, equation (1) becomes Jensen-Shannon divergence (JSD). This modified CJSD is useful when distributions are similar and it is hard to distinguish them. A further modification was proposed in [39] which extends CJSDs to exploit the metric properties of JSD. Metric-Chisini-Jensen-Shannon divergences (M-CJSD) is defined as:

$$M - CJSD(P||Q) = \sqrt{CJSD(P||Q)} \tag{2}$$

Some M-CJSD based kernels are:

$$K_{Amplified} = M - CJSD(P||Q)e^{-\frac{|x_i - x_j|^2}{2\sigma^2}} \tag{3}$$

$$K_{Scaled} = e^{-M - CJSD(P||Q)\frac{|x_i - x_j|^2}{2\sigma^2}} \tag{4}$$

$$K_{Amplified-Scaled} = M - CJSD(P||Q)e^{-M - CJSD(P||Q)\frac{|x_i - x_j|^2}{2\sigma^2}} \tag{5}$$

Likewise, replacing M-CJSD with CJSD in (3), (4) and (5), we get CJSD based kernels. In this paper, we study these kernels for Arithmetic, Geometric, and Harmonic (AM, GM, HM) means. Therefore, we employ AM, GM, HM for Amplified, Scaled, and Amplified-Scaled kernel versions for both M-CJSD and CJSD. This results in a total of 18 kernels.

## 4 Gesture Myo Data

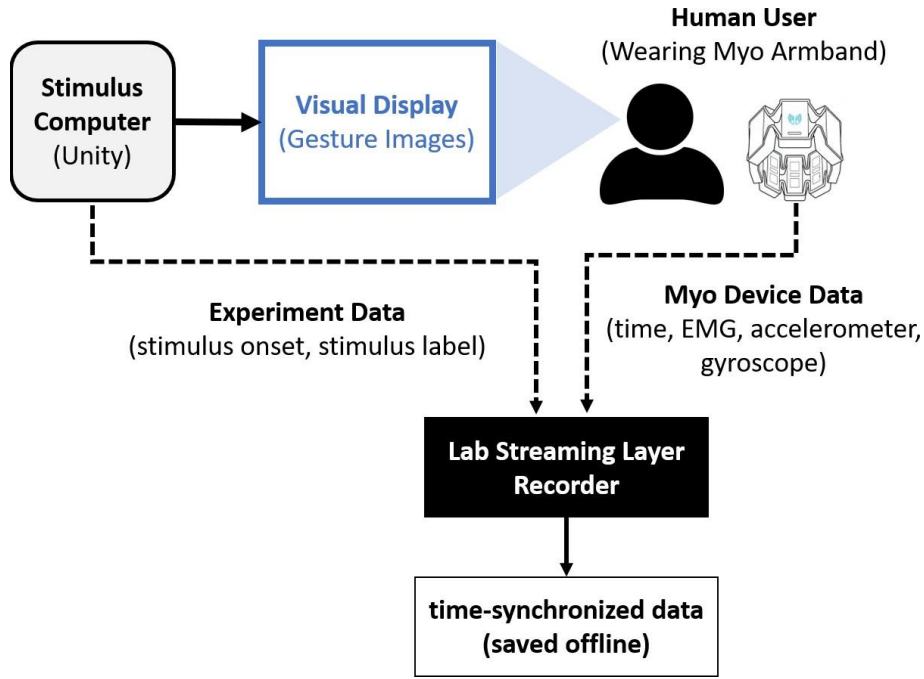

Fig. 1: Experimental Design.

As we set out to investigate how soldiers' can increase situational awareness by using gestures in military operations, our data came from physiological experimentation, however, this paper is focused on use by soldiers versus medical doctors [8]. Gesture datasets were collected by a Myo armband, made by *Thalamic Labs* (Figure 2). The device contains an *inertial measurement unit (IMU)* and *electromyography (EMG)* sensors to measure changes along the up-down, left-right, and front-back axes and changes in muscle activity, respectively. These measurements are based on electrical stimulation created by flexing forearm muscles coupled with a 9-axis IMU, which in fact is comprised of 3-axis gyroscope, a 3-axis accelerometer, and a 3-axis magnetometer to track arm movement.

Table 1: Data from all sensors in the experiment

| Sensor Name | Channel |
|---|---|
| Electromyogram (EMG) | $1 - 8$ |
| Accelerometer | $9 - 11$ |
| Gyroscope | $12 - 14$ |
| Myo Pose Classification | 15 |
| Class Label (gesturing or not gesturing) | 16 |

The goal was to develop an algorithm that can automatically provide the likely start and stop indices of each gesture segment enabling the parsing of (and labeling of) gesture data from the actual experiment for further analysis and modeling. Subjects wore the Myo armband on their left arm and performed a series of NATO gestures one after the next according to a display on a screen in front of them. The display was run in the Unity game engine. Subjects relaxed their arm in between gestures. Timing information of the visual display and data from the Myo armband were recorded and synchronized through custom code using LabStreamingLayer (LSL) [17]. This synchronization produced a data frame (Table 1), that enabled analyses of all data within the same time-space (Figure 1). Detailed description of the experimental design is given in [21].

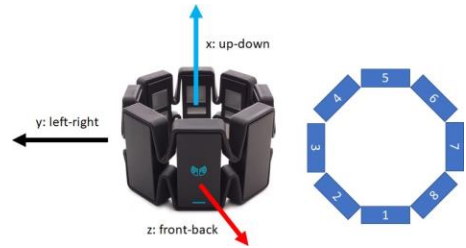

Fig. 2: The Myo armband used in actual experimentation with associated IMU axes and EMG sensor positions.

## 5    Signal-Image Encoding

We consolidated 3-axis accelerometer, 3-axis gyroscope from IMU, and 8-axis EMG to produce a fused signal of length 14. After preprocessing, our data consisted of a total 38507 instances, with 24845 instances of no gesture, and 13662 instances of gestures. Each instance represented a signal, derived from the 3 different sensors (Accelerometer, Gyroscope, EMG). This resulted in a channel of length 14. In our experiment, we used the channel at each axis as a feature. Therefore, our data instances were vector valued in 14 dimensions.

Because we are dealing with short signals of length 14 each, in order to produce an image of size 4×4, we make it of length 16 by adding 2 zeros at the end. This is called *Zero Padding* which is useful in many applications allowing us to increase the frequency resolution arbitrarily. This is like treating a signal as if the short burst is followed by silence and does not impact our data significantly. However, for a long zero padding a reasonable approximation of the actual note is required. Next, we scale each signal between 0-255, and then use Python Imaging Library (PIL, aka Pillow) library to save encoded signal as an image.

In order to compute texture features,we use GIST descriptor which uses a wavelet decomposition of an image [28]. After preprocessing, our data consisted of a total 36140 samples with 23043 no gesture, and 13097 gestures. An image has 1 GIST descriptor of 512 dimensions. Therefore, our data consisted of a total 36140 instances, each representing a single GIST descriptor of 512 dimensions. Data thus obtained poses challenges with high dimensionality (Figure 4).

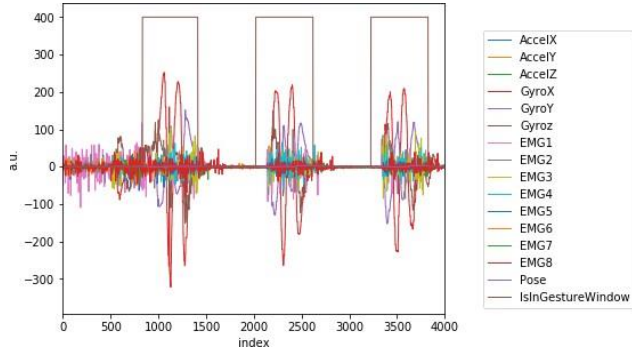

Fig. 3: Illustration of data originated from the Myo device and Unity gesture experiment program. Square wave wraps cleanly around the rest of the data, shown as noisier channels within the square wave.

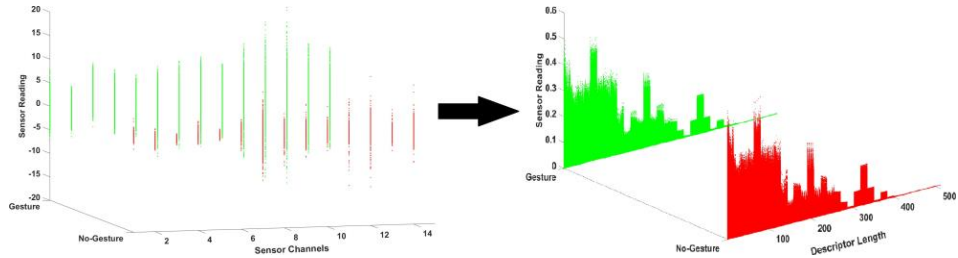

Fig. 4: (Left) Original signal data (Right) Image data after transforming signal.

## 6 Experiments

In order to detect if a gesture was made, we explore classical machine learning, deep learning, transfer learning, and anomaly (novelty) detection approaches mentioned in Section 1. We validate applied methods on fused sensor data (which we call *signal data*), on images obtained with our *signal-image* encoding scheme, and on GIST features (Section 5). For all of our experiments, we build a bi- nary classification problem by doing 80:20 percent data split into train and test/validation datasets, and label data instances as *Gesture* and *No-Gesture*.

Table 2: Model summary of trainable parameters

| Layer (type) | Output Shape | Param # |
|---|---|---|
| resnet50 (Model) | (*None*, 8, 8, 2048) | 23587712 |
| atten_1 (Flatten) | (*None*, 131072) | 0 |
| dense_1 (Dense) | (*None*, 1024) | 134218752 |
| dropout_1 (Dropout) | (*None*, 1024) | 0 |
| dense_2 (Dense) | (*None*, 2) | 2050 |

Total params:      157, 808, 514
Trainable params:      135, 275, 522
Non-trainable params: 22, 532, 992

For the initial set of experiments, we employ transfer learning approach. This requires selection of a suitable a pre-trained model. For this purpose, we tried different pre-trained models from the list of available models for image classification with weights trained on *ImageNet* (Xception, VGG16, VGG19, ResNet, ResNetV2, InceptionV3, etc.) [9]. Our experiments show that all models give similar classification results, therefore, we report results only for ResNet50. A summary of the trainable parameters of the model is provided in Table 2. We ne-tune the model by tweaking its parameters for our data and freeze all of the layers except the last 4 convolutional layers which we use for training. Fine-tuning avoids limitations of model by not training from scratch on small data and saving training time (because less parameters will be updated in training). We evaluate model performance by computing validation accuracy and validation loss.

For the next set of experiments, we explore the performance of CJSD and M-CJSD kernels in SVM classification (mentioned in Section 2), and that of DNN on both, the signal data and the GIST features. We estimated probability densities of CJSDs and M-CJSDs using multivariate kernel density estimation (KDE), a nonparametric approach, on each dataset [34,31,44]. This distribution is defined by a smoothing function and a bandwidth value that controls the smoothness of the resulting density curve. Results were validated through 10-fold nested cross-validation on the randomized datasets constructed from binary classification problem. It is important to note that datasets used in experimentsfor each type of CJSD and M-CJSD kernel (A.M., G.M., H.M.) were randomized using unique random number seeds. For fair comparison, we tested RBF kernels with the same randomized datasets. This resulted in a total of 21 kernel versions(3 for RBF and 18 explained in (Section 3).

For DNN hyperparameter tuning, we searched over a large grid with *Batch Size, Epochs, Optimizer, Learn Rate, Momentum, Initial Mode, Activation, Dropout Rate, Weight Constraint,* and *Neurons*. DNN and radial basis function (RBF) kernel are used as the performance benchmark for comparison with CJSD and M-CJSD kernels. We report the classification results with sample mean and standard error in error bar plots (see Section 7).

For GIST data, we employed PCA and used Pareto charts to select the top *k* principal components that explained over 99% variance [16]. This resulted in a dimensionality of 20. For visualization, we employed t-SNE [24] (Figure 5).

Our goal is to increase confidence in decision-making in military operations using our model's ability to detect gestures correctly for soldiers being able to decide whether a new observation belongs to the same distribution as existing observations (inlier), or should be considered as different (outlier). Therefore, for the final set of experiments, we prepare our training data which is not polluted by outliers and detect whether a new observation is an outlier. In this context an outlier is also called a novelty. We perform anomaly detection by training the model only on the positive (known) class dataset and predicting negative (unseen) classes. Out of many available algorithms for anomaly detection, we choose One-class SVMs [6], Isolation Forests [23] and Gaussian Mixture Model (GMM) [33]. We use isotonic regression to convert the output of the GMM to a probability score. We train our models on No-Gesture as a positive class and use it to predict Gesture as a negative class. We evaluate models' performances by comparing their respective sensitivity (true positive rate) and specificity (true negative rate) along with confusion matrices.

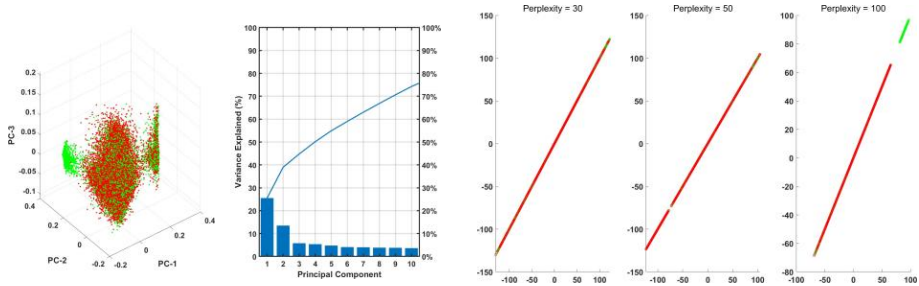

Fig. 5: (Left) PCA scatter and pareto plots. Notice the thick blue line corresponding to explained variance of 76% for 10 principal components. (Right) t-SNE scatter subplot. Best seen in color.

We used LIBSVM library in MATLAB to employ aforementioned kernels [5]. For all other experiments and data preprocessing, we used the latest versions of Python and Keras framework with TensorFlow as a backend [7]. Our computing system consisted of 128 GB RAM for CPU, and NVIDIA Quadro P3200 6144MB - Memory Type: GDDR5 (Samsung). Performance in terms of time taken to run by each model is reported in Table 3.

## 7    Results

### 7.1    Classification

*Transfer Learning* achieved 64.52% average accuracy with 4.84 loss on validation set (Figure 6). For signal data, kernels and DNN achieved classification accuracy ranging between 70%-88%. For GIST data, it ranges between 63.76%-67.60%. Among all tested methods, each except *scaled_AM* gave higher classification ac-curacy on signal data in comparison to GIST data (67%) with statistical significance at the 95% confidence level (Figure 7). We use the overlap in confidence in-tervals to check the statistical significance. As the intervals do not overlap, there is at least 95% confidence (with $p$-value at the $p < 0.05$ level of significance).

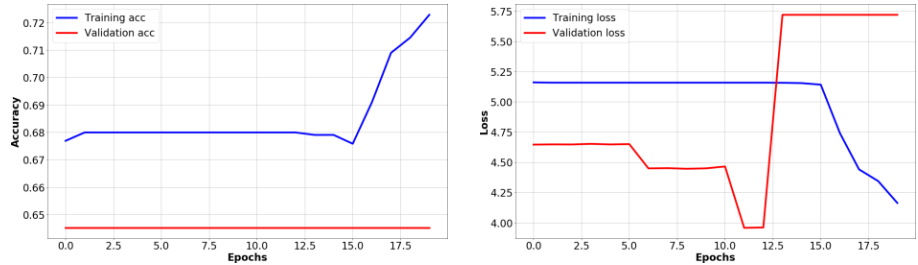

Fig. 6: Fine tuning with weights trained on ResNet50 model by freezing all the layers except the last 4 layers we obtain (Left) Training and Validation Accuracy (Right) Training and Validation Loss.

On the other hand, run-time comparison shows that CJSD and M-CJSD based kernels outperform DNN significantly. Because we tested 21 kernels in total, approximate time taken by each kernel is between 20-30 minutes for both GIST and signal datasets. Time can be computed by dividing total time with 21 from Table 3.

Table 3: Comparison of models' performance

| Model | Data | Platform | Time taken to run the model (Hours-Minutes-Seconds) |
|---|---|---|---|
| Transfer Learning | Image | GPU | 1 : 24 : 37 |
| DNN | GIST | GPU | 3 : 45 : 07 |
| DNN | Signal | GPU | 3 : 53 : 06 |
| CJSD, M-CJSD, RBF Kernels in SVM | GIST | CPU | 9 : 57 : 00 |
| CJSD, M-CJSD, RBF Kernels in SVM | Signal | CPU | 7 : 18 : 00 |
| Anomaly Detection (One-class SVM, Isolation Forest, GMM) | Signal and GIST | CPU | < 25 seconds |

## 7.2   Novelty Detection

Table 4 summarize results from anomaly detection approaches described in Section 6. Models were trained on Gesture instances. Among all tested methods, for signal data, *Isolation Forest* achieves the highest accuracy of 90.51% with 100% sensitivity (blue). Moreover, for both signal and GIST datasets, GMM with Isotonic Regression achieves very similar results with respective accuracy of 90.09% and 89.79% and 100% sensitivity (green). Thus, with our proposed signal-image encoding approach; we were able to perform detection task while preserving most of the information in data transformation. It is also to significant that *GMM* was able to achieve higher accuracy over kernels and DNN. Figure 8 provides corresponding confusion matrices. For signal data, notice that *Isolation Forest* was able to detect all 24844 *Gesture* instance correctly with only 1 misclassification. On the other hand, for both signal and GIST datasets, *GMM* with *Isotonic Regression* was able to detect all 24845 and 23043 *Gesture* instances correctly with 0 misclassifications, however, with comparatively low false positives. Also, notice that anomaly detection methods took comparatively much less run time (of the order of a few seconds) than did other classical and deep learning methods (Table 3).

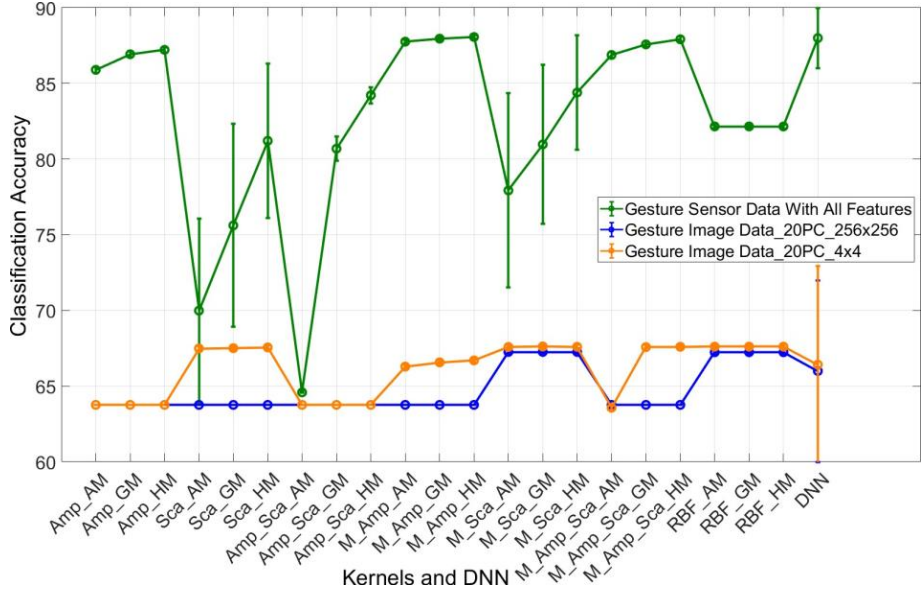

**Fig. 7:** Error bars show accuracy achieved by CJSD, M-CJSD, RBF kernels in SVM classification and DNN for sensor (green), GIST with image resized to 256 x 256 (blue), and GIST with original 4 x 4 image (orange). Results are reported for 3 randomized data versions (*AM, GM, HM*) for CJSDs and M-CJSDs. For signal data (green), notice that DNN and all metric CJSD kernel version except its scaled version outperform other kernels achieving around 88% average accuracy. RBF performs the same for each randomized data version achieving around 82% average accuracy. For GIST features (orange and blue), the highest accuracy achieved by some kernels is around 67% and for DNN it is around 66%. Finally, notice the high standard error for DNN and some kernels.

**Table 4:** Anomaly detection results after training on *Gesture* instances as explained in figure 8. Best model is colored in blue.

|  | Accuracy | Sensitivity | Specificity |
|---|---|---|---|
|  | On sensor fusion (signal) data model | | |
| One Class SVM | 5.96% | 1.02% | 50.86% |
| Isolation Forest | 90.51% | 100% | 4.32% |
| GMM with Isotonic Regression | 90.09% | 100% | 0% |
|  | Using GIST feature descriptors | | |
| One Class SVM | 44.64% | 44.11% | 49.27% |
| Isolation Forest | 86.91% | 96.51% | 2.44% |
| GMM with Isotonic Regression | 89.79% | 100% | 0% |

Fig. 8: (Left to Right) Confusion matrices for *One Class SVM, Isolation Forest* and *GMM* with *Isotonic Regression* trained on 80% Gesture instances, and tested on 20% *Gesture + 100% No-Gesture* instances with features extracted from (Above) fusion sensor (signal) data (Below) GIST Feature descriptor.

## 8    Conclusion

In this paper we made several contributions concerning the challenges to analyze signal data originating from multiple sources. In particular, we adopted a *multi-sensor data fusion* approach using IMU and EMG sensors to leverage abundant, raw, digital information for strategic advantage. Impetus behind our research arise from growing consideration of potential use of *Internet of Things (IoT)* devices in military's *Multi Domain operations (MDO)*.

Massive data generated from IoT devices allows the full realization of system to provide situational awareness. In order to leverage communication in C2 (Command and Control) systems, and limitations of *Blockchain* technology which is commonly used to facilitate secure sharing of IoT datasets, we introduced a *signal-image encoding* approach to modify information (signal) by transforming it to an image which is invertible and easier to visualize. We proposed *zero padding* to address the challenges of very short signals in conversion. Significance of multi-sensor fusion and signal-image encoding approaches is to transform senseless and unstructured data into structured data from which meaningful information can be extracted to be used as actionable insights that enable intelligent military *decision-making*. In general, our approach is applicable to military environments where IoT devices maybe used.

Moreover, we investigated the challenge of enabling an intelligent identification and detection operation and demonstrated the feasibility of the proposed *Machine Learning, Deep Learning and Anomaly Detection* models to support a detection and identification of hand gestures. We evaluated all methods on hand gesture datasets before and after signal-image encoding, and extracted GIST descriptors from images. Results from anomaly detection with *GMM + IsotonicRegression* confirmed that it achieved statistically significantly similar accuracy for both, signal and GIST datasets. Thus, our proposed encoding approach wasable to perform detection task while preserving most of the information in datatransformation. It is also showed that *GMM* was able to achieve higher accuracy over classical and deep learning methods. Another virtue of tested anomaly detection methods is associated extremely low computational complexity in terms of time and memory over other classification approaches (Table 3).

Future work will extend our analysis to a more complex heterogeneous IoT environment. We will investigate the impact of governing parameters (IoT topology, connectivity, security, etc.) on model performance for data fusion. In this work we addressed the challenges of signal-image encoding arising from very short signals. We will explore methods to encode short signals for secure information transformation via wireless network to support decision making.